\documentclass[apj]{emulateapj}
\usepackage{float}
\usepackage{apjfonts}
\usepackage{amsmath}

\usepackage{amsmath,amssymb}

\slugcomment{Received 2007 May 14; accepted 2007 May 29; published 2007 June
  27}
\shorttitle{MAGNETIC BUOYANCY AND COHERENT STRUCTURES}
\shortauthors{KERSAL\'E, HUGHES, \& TOBIAS}

\begin{document}
\bibliographystyle{apj}

\title{The nonlinear evolution of instabilities driven by magnetic buoyancy: A
  new mechanism for the formation of coherent magnetic structures}

\author{Evy Kersal\'e, David W. Hughes, and Steven M. Tobias}
\affil{Department of Applied Mathematics, University of Leeds,
  Leeds LS2 9JT, U.K.}
\email{kersale@maths.leeds.ac.uk}

\begin{abstract}
  Motivated by the problem of the formation of active regions from a
  deep-seated solar magnetic field, we consider the nonlinear
  three-dimensional evolution of magnetic buoyancy instabilities resulting
  from a smoothly stratified horizontal magnetic field. By exploring the case
  for which the instability is continuously driven we have identified a new
  mechanism for the formation of concentrations of magnetic flux.
\end{abstract}

\keywords{Sun: interior --- Sun: magnetic fields --- instabilities --- MHD}

\section{Introduction}

\label{sec:intr}

Solar active regions are the surface manifestations of a deep-seated,
predominantly toroidal magnetic field. Although it is generally agreed that
the solar magnetic field is maintained by some sort of hydromagnetic dynamo,
the mechanism by which this is effected is far from understood. Consequently,
the strength, structure, and location of the interior magnetic field are
similarly unknown. However, most recent solar dynamo models, despite their
significant differences in other respects, postulate that the tachocline, the
thin layer of strong velocity shear located at the base of the convection
zone, plays an important role in the generation of toroidal field by the
shearing of a weaker poloidal component \citep[see, e.g., the review
by][]{smt2007}. Given this premise, it is important to address the nature of
the initial escape of the magnetic field from the tachocline, its subsequent
ascent through the convection zone, and its eventual emergence at the
photosphere.

Owing to the vast range of scales across the convection zone, it is impossible
to model realistically all of these stages in one calculation. Here we
concentrate solely on the instability of a layer of magnetic field, with the
aim of clarifying the physics of the formation of coherent magnetic structures
from a much larger scale field.

A vertically stratified, horizontal magnetic field can be unstable to magnetic
buoyancy instability provided that the field decreases sufficiently rapidly
with height. It is important to note the non-trivial distinction between
instability to two-dimensional modes in which the field lines remain straight
(interchange modes) and instability to fully three-dimensional modes. The
former are essentially destabilized by a decrease with height of $B / \rho$
(where $B$ is the magnetic field strength and $\rho$ the density), the latter
simply by a decrease with height of $B$. The physics underlying this
difference is elucidated in \citet{1987GApFD..39...65H}.

The linear theory of magnetic buoyancy instabilities has been intensively
studied over a number of years \citep[e.g.][]{Newc61, 1966ApJ...145..811P,
  1970ApJ...162.1019G, 1978RSPTA.289..459A}. The nonlinear development of the
instability, particularly with respect to a deep-seated solar field, has,
inevitably, received less attention. \citet{1988JFM...196..323C} investigated
the nonlinear evolution of interchange modes resulting from a slab of uniform
horizontal magnetic field embedded in a convectively stable and otherwise
field-free atmosphere. Here the instability is driven by a density jump at the
upper interface of the magnetic slab --- an extreme form of magnetic buoyancy
instability. This interfacial instability generates a strong shear flow that
leads, by a secondary Kelvin-Helmholtz instability, to the formation of strong
vortices. The subsequent nonlinear evolution is then governed by pairwise
vortex interactions, which can even act so as to drag down pockets of strong
magnetic field. The three-dimensional instability and nonlinear evolution of
the same basic state was examined by \citet{1995ApJ...448..938M} and
\citet{2000MNRAS.318..501W}. The initial evolution again takes the form of
interchange modes, leading to the formation of vortex tubes. However,
neighbouring vortex tubes of opposite sign are unstable to a longitudinal
instability \citep{crow70}; this in turn causes the magnetic field to adopt an
arched structure. By contrast, \citet{2001ApJ...546..509F} investigated the
nonlinear evolution of a smoothly varying magnetic field, with a profile
chosen such that the initial state is unstable to three-dimensional modes, but
stable to interchange modes. She shows the formation of arched magnetic
structures, which maintain a reasonable degree of coherence as they rise. A
recent review of the implications of magnetic buoyancy instabilities for the
solar tachocline is given by \citet{dwh2007}. Related simulations of magnetic
buoyancy in sub-surface regions have also been performed
\citep[e.g.][]{2004A&A...426.1047A, 2005Natur.434..478I}.

Inspection of active regions on the solar surface suggests the emergence
through the photosphere of a buckled, toroidal magnetic field. However, very
little is known about the structure, strength and orientation of the
sub-surface field. For example, which properties of the observed surface field
result from the initial instability and which result from later interactions
in the convection zone? Our aim in this Letter is to seek further
understanding of the instability by examining the nonlinear evolution from a
very simple equilibrium state. We consider an atmosphere that is in both
magnetohydrostatic and thermal equilibrium and that has a linear magnetic
field profile. We concentrate on modes that are intrinsically
three-dimensional; i.e.\ the field gradients considered are such that the
basic state is unstable to three-dimensional modes but stable to interchanges.

Furthermore, we impose boundary conditions such that the source of the
instability is maintained for all times, thus allowing us to consider the
long-term evolution. This is an important difference from the earlier works
cited above, all of which consider ``run-down'' experiments, in which the
potential energy stored in the initial field configuration is rapidly
converted into kinetic energy and which is then slowly dissipated.

We consider four cases, characterised by two different field strengths and two
different choices of initial conditions. We have identified a completely new
mechanism for the formation of flux concentrations; this is of potential
significance in the solar context in which the formation of localised regions
of strong field from a weaker larger scale field is a crucial component of the
flux emergence process.

\begin{figure}
  \epsscale{0.9}
  \plotone{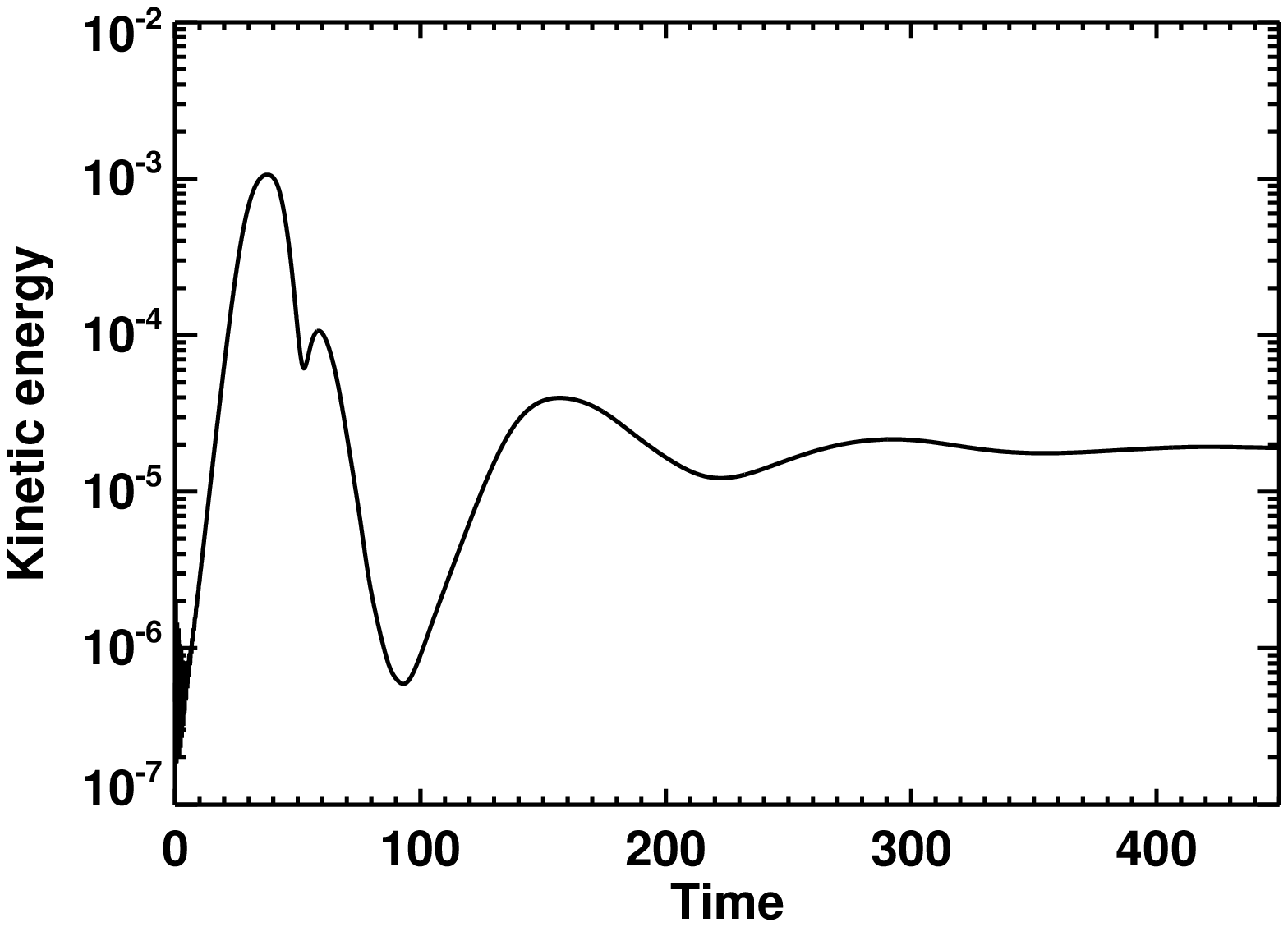}
  \plotone{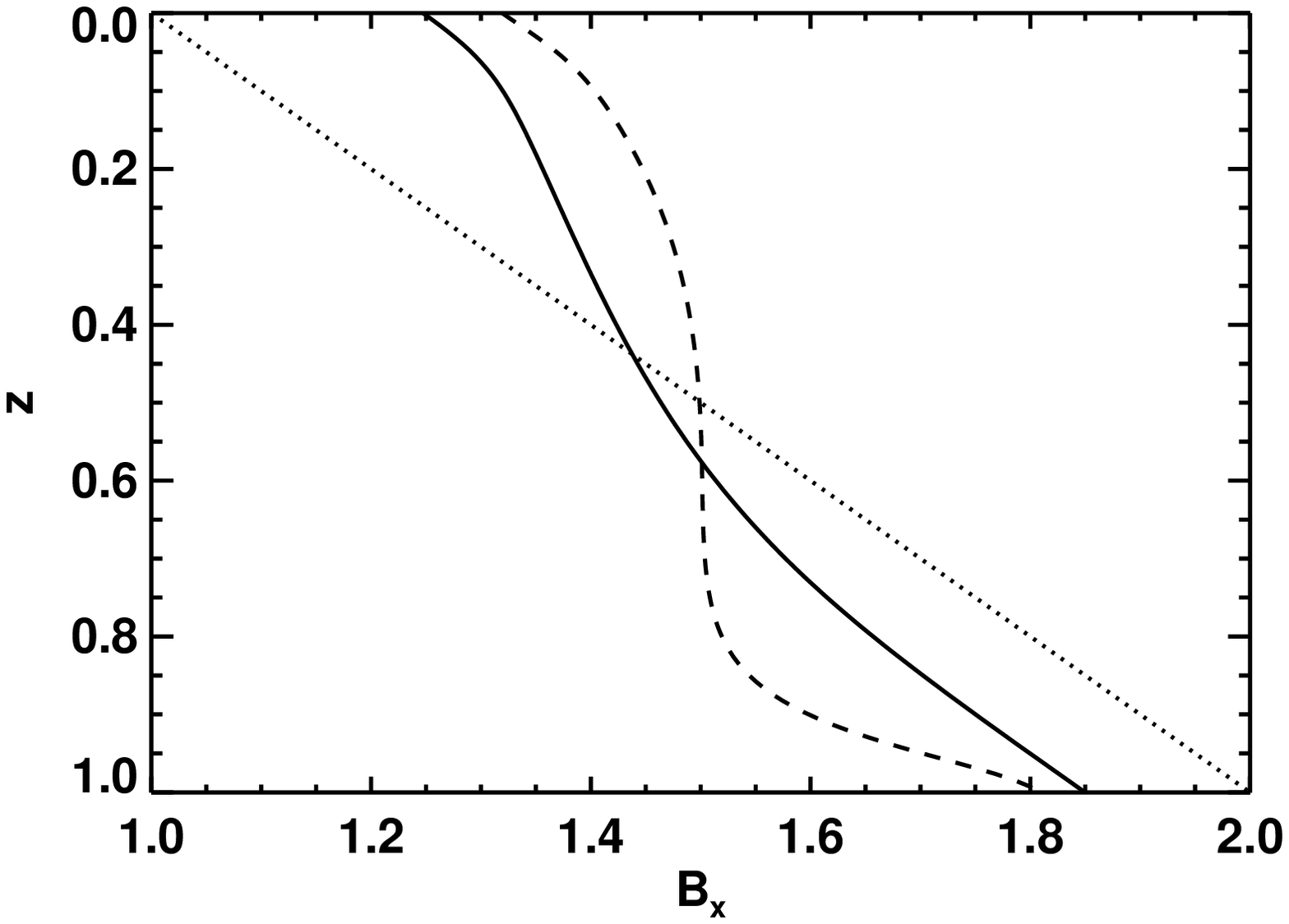}
  \figcaption{\label{fig:f01} Model with $\alpha=\frac{1}{4}$ and eigenfunction
    perturbation. ({\sl a}) Evolution of the kinetic energy density. ({\sl b})
    Horizontal average of the magnetic field $B_x$ vs. depth, initially
    ({\sl dotted line}), at $t \approx 50$ ({\sl dashed line}), and in the
    final steady state ({\sl solid line}).}
\end{figure}

\section{Mathematical formulation and parameters}

We consider a Cartesian layer of perfect gas governed by the equations of
three-dimensional compressible MHD. The dimensionless representation of these
equations is obtained by scaling lengths with the layer depth, time with the
isothermal sound crossing time at the top surface ($z=0$), and temperature,
density, and magnetic field with their respective values at the top surface.
The physical properties of the fluid are entirely determined by four
dimensionless numbers: the Prandtl number $P_r$, the ratio of viscous to
thermal diffusivity; $\zeta$, the ratio of magnetic to thermal diffusivity;
$\gamma$, the ratio of specific heats; and $C_k$, the nondimensional thermal
conductivity.

The basic state that we consider is an exact equilibrium, dependent only on
depth $z$, with a unidirectional horizontal magnetic field $B_x(z) = 1 +
z/H_b$. The density and temperature are determined from solution of the
magnetohydrostatic equations. In the absence of a magnetic field, the
atmosphere takes the form of a plane polytrope, with polytropic index $m$ and
constant temperature gradient $\theta$. The strength of the magnetic field is
changed by varying $\alpha = 2 / \beta \propto B^2$, where the plasma-$\beta$
is the ratio of the thermal to the magnetic pressure. A more detailed
description of the governing equations and parameters is contained in
\citet{1998ApJ...502L.177T}.

We consider perturbations to this basic state, subject to the following
boundary conditions: all variables are taken to be periodic in both horizontal
directions; the top and bottom boundaries are stress-free, impermeable, and
isothermal. For the magnetic field we choose to maintain a fixed gradient of
$B_x$ ($\partial B_x / \partial z = 1/H_b$) at $z=0, 1$, together with
$\partial B_y / \partial z = 0$ and $B_z = 0$. Our aim in this Letter is to
examine the nonlinear development of modes that are intrinsically
three-dimensional. By an extensive linear analysis we have identified the
regions in parameter space where three-dimensional modes ($k_x \ne 0$, $k_y
\ne 0$) are unstable but interchange modes ($k_x = 0$) are stable. In the next
section we discuss the nonlinear evolution resulting from two choices of
parameters and two different types of perturbation to the basic state.

\section{Nonlinear evolution of three-dimensional modes}

The nonlinear evolution of three-dimensional magnetic buoyancy instabilities
is studied by solving numerically the equations of compressible MHD with a
modified version of the hybrid finite-difference-pseudospectral parallel code
used by \citet{1996ApJ...473..494B} (hydrodynamic) and
\citet{1998ApJ...502L.177T} (MHD). We consider the evolution from two basic
states, distinguished by the value of the field strength; the other parameters
are held fixed at $P_r = \zeta = 2\times 10^{-2}$, $C_k = 2.5 \times 10^{-2}$,
$\theta = 2$, $m=1.6$, $H_b=1$, and $\gamma = 5/3$. Furthermore, we consider
two different perturbations to each basic state: one consisting of the most
unstable eigenfunction, the other of random noise. In each case, the
horizontal dimensions of the Cartesian computational box are determined by the
wavenumbers $k_x$ and $k_y$ of the most unstable mode such that one wavelength
fits in the $x$-direction, along the initial magnetic field, and four
wavelengths fit transverse to the initial field, in the $y$-direction.

\subsection{Evolution to a Steady State}
\label{etass}

\begin{figure}
  \epsscale{0.9}
  \plotone{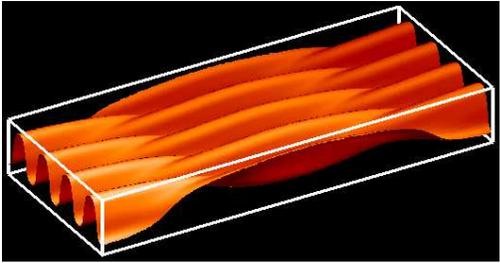}
  \figcaption{\label{fig:f02} Isosurface of magnetic energy $(B^2/2=1)$ at
  $t \approx 45$ for the same model as in Fig.~\ref{fig:f01}.}
\end{figure}

Here we consider a basic state with $\alpha = 1/4$ perturbed by the most
unstable eigenfunction, namely, that with growth rate $\sigma = 0.16$ and with
wavenumbers $k_x = 0.87$ and $k_y = 8.38$ (hence, the dimensions of the
computational domain are $L_x = 7.24$ and $L_y=3$). The evolution is
essentially linear up to time $t \approx 21$. For $21 \lesssim t \lesssim 37$
the kinetic energy continues to grow (see Fig.~\ref{fig:f01}{\sl a}). During
this stage, broad upflows carry strong magnetic fields from the bottom of the
layer, while narrow plumes pull down weaker field, leading to the formation of
arched structures, as shown in Figure~\ref{fig:f02}. Figure~\ref{fig:f01}{\sl
  b} shows the mean result of this nonlinear evolution, with a significant
reduction of the gradient of the horizontally averaged magnetic field $B_x$.
Initially constant throughout the layer ({\sl dotted line}), the vertical
gradient of $B_x$ becomes nearly zero in a large fraction of the domain at $t
\approx 50$ ({\sl dashed line}). The nonlinear reorganization has therefore
acted to remove the driving mechanism of the instability. There is an obvious
analogy with thermal convection, in which saturation is achieved by a
nonlinear reorganization leading to an adiabatic core with thin thermal
boundary layers.

The arched configuration shown in Figure~\ref{fig:f02} is however transient;
following this initial nonlinear reorganization the system relaxes to a state
with weak flows and a magnetic field profile that is close to critical. As can
be seen from Figure~\ref{fig:f01}{\sl b}, the field gradient in the final
state is reduced everywhere in the layer compared with its initial value. This
subsequent evolution is permitted by the choice of boundary conditions, in
which the gradient of $B_x$ (but not the field itself) is fixed at the
boundaries.

\subsection{Evolution to Time-dependent Concentrated Flux States}

\begin{figure}
  \epsscale{0.99}
  \plotone{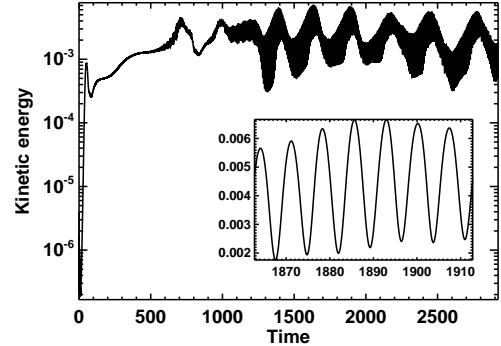}
  \plotone{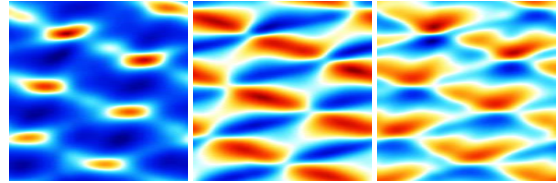}
  \figcaption{\label{fig:f03} Model with $\alpha = \frac{1}{4}$ and random
    perturbation. ({\sl a}) Evolution of the kinetic energy density. ({\sl b})
    Space-time plots, with $y$ horizontal and $t$ vertical, of magnetic energy
    $(1.17 \lesssim B^2/2 \lesssim 1.78)$, $|v| \lesssim 0.05$, and $|w|
    \lesssim 0.06$; $ 1863 \lesssim t \lesssim 1913$; $z\approx 0.7$. }
\end{figure}

\begin{figure}
  \epsscale{0.9}
  \plotone{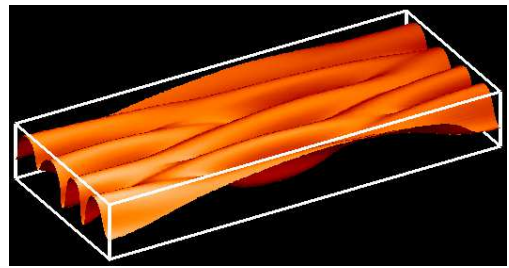}
  \figcaption{\label{fig:f04} Isosurface of magnetic energy $(B^2/2=1)$ at
   $t \approx 74$ for the same model as in Fig.~\ref{fig:f03}.}
\end{figure}

First we consider the evolution from the same basic state as in \S\ref{etass}
(i.e.\ with $\alpha = 1/4$), but with the perturbation taking the form of
small-amplitude random noise. As shown in Figure~\ref{fig:f03}{\sl a}, the
linear evolution (up to $t \approx 36$) and the initial nonlinear phase, as
measured by the temporal evolution of gross properties such as kinetic energy,
are similar to those portrayed in Figure~\ref{fig:f01}{\sl a }. The first
nonlinear reorganization again leads to the formation of arched structures, as
shown in Figure~\ref{fig:f04}, although, as a result of the initial
conditions, these are less regular than those of Figure~\ref{fig:f02}.
Crucially, these irregularities play a significant role in modifying the
subsequent behaviour. From $100 \lesssim t \lesssim 450$ the kinetic energy
increases monotonically until the onset of a secondary oscillatory
instability. From $t\approx 1200$ the system evolves on two disparate
timescales: a short cycle with period $\approx 7$ modulated on larger
timescales $250 \lesssim t_\text{mod} \lesssim 300$.

The underlying mechanism of the short period oscillations can be understood by
inspection of Figure~\ref{fig:f03}{\sl b}, which shows the temporal evolution
for fixed values of $x$ and $z$ of the magnetic energy, the transverse
horizontal velocity $v$, and the vertical velocity $w$. The left panel shows
the periodic formation of a concentration of magnetic energy, which drifts
slowly with velocity $0.03$. The central and right panels indicate that the
concentrations of magnetic energy result from convergent downflows associated
with rolls in the $yz$-plane. The magnetic flux becomes concentrated between
the two counter-rotating rolls; it thus becomes buoyant and rises rapidly.
This drives a countercell, which diverges at the flux concentration and
thereby destroys it. After some reorganization, the initial cellular flow is
reestablished but displaced relative to its initial position. Reformed
downflows again lead to a concentration of flux and the entire process is
repeated.

\begin{figure}
  \epsscale{1.}
  \plotone{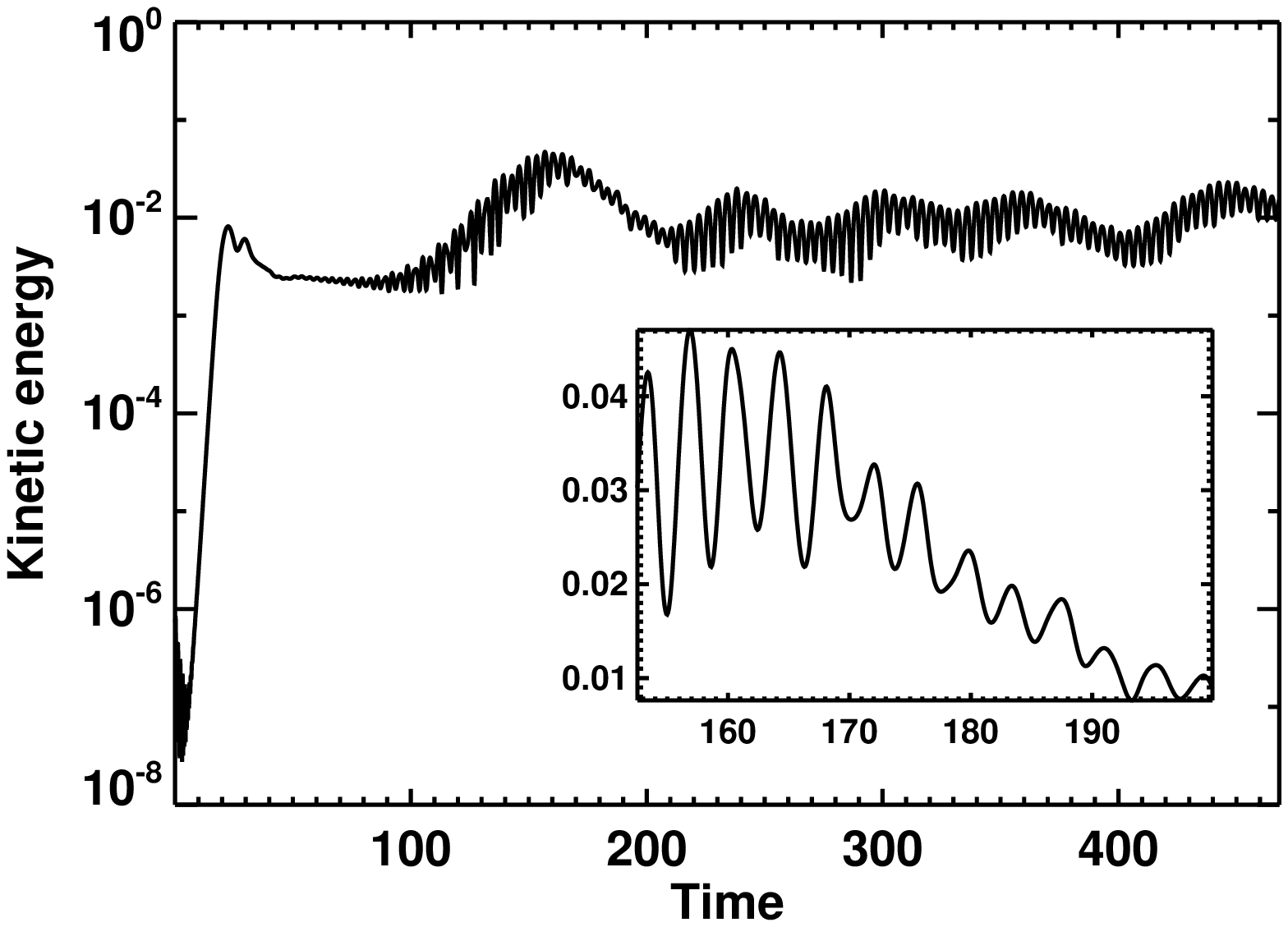}
  \plotone{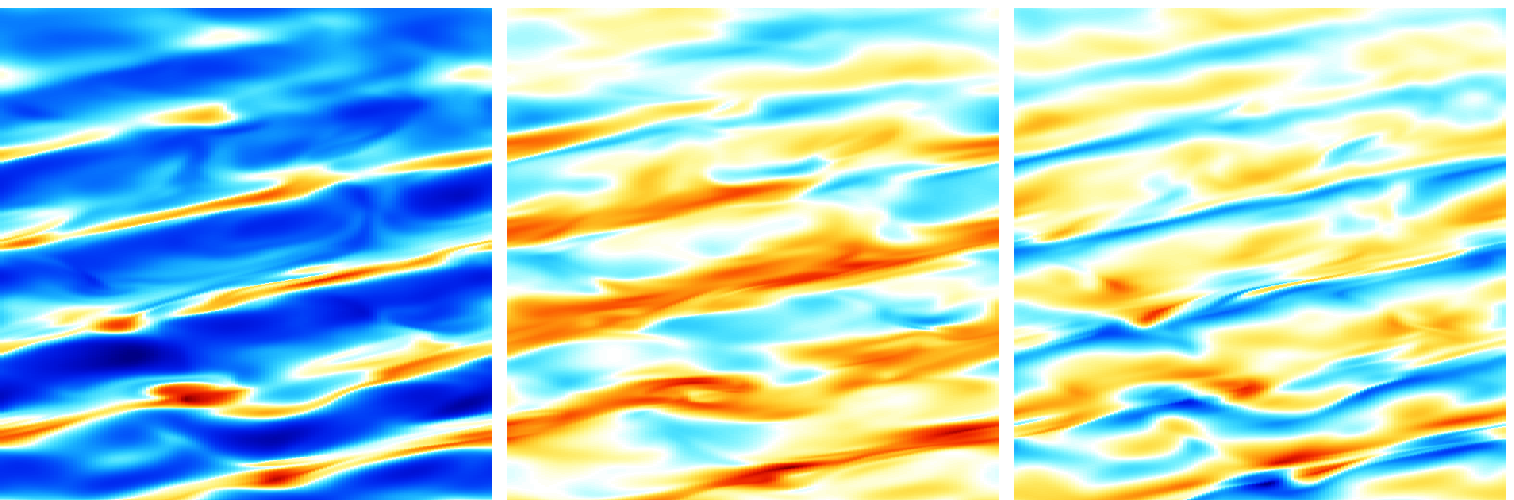}
  \figcaption{\label{fig:f05} Same as Fig.~\ref{fig:f03}, but with $\alpha
    =1$. In panel {\sl b} $0.65 \lesssim B^2/2 \lesssim 2.27$, $|v| \lesssim
    0.51$, $|w| \lesssim 0.56$; $153 \lesssim t \lesssim 200$; and $z \approx
    0.5$.}
\end{figure}

\begin{figure}
  \epsscale{0.9}
  \plotone{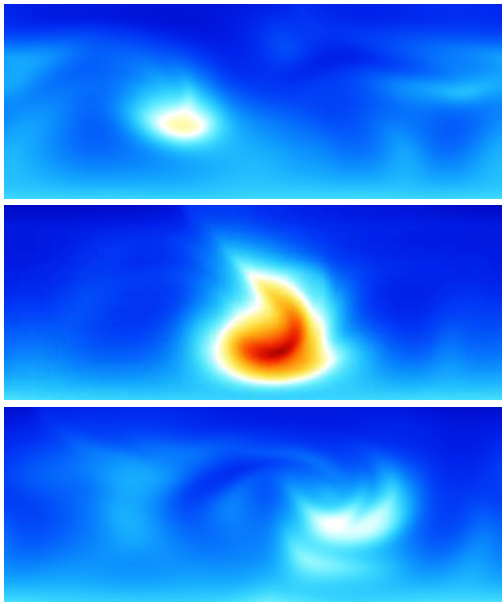}
  \figcaption{\label{fig:f06} Slices in the $yz$-plane of magnetic energy at
    times $t \approx 152.9$, $154.7$, $156.5$. The peak magnetic energy
    $\approx 3$.}
\end{figure}

Next we examine the evolution from an equilibrium with a stronger field
($\alpha =1$). For this case the evolution is qualitatively similar regardless
of the nature of the initial perturbation; here we shall describe the case in
which the perturbation takes the form of random noise. The most unstable mode
has wavenumbers $k_x = 0.86$ and $k_y = 9.75$, and growth rate $\sigma =
0.45$. Figure~\ref{fig:f05}{\sl a} shows that the temporal evolution is
broadly similar to that of the weaker field case portrayed in
Figure~\ref{fig:f03}{\sl a}. The secondary instability is evident, leading to
a modulated periodic state. The increase in initial field strength leads, as
expected, to a more vigorous instability (higher growth rate and increased
saturation level) and to a shorter period oscillation of the nonlinear state.
As in the example above, flux concentration occurs via converging downflows.
The peak field in the flux concentrations is significantly stronger than the
average initial field. In this case, as can be seen in Figure~\ref{fig:f06},
the flux concentrations are never completely dispersed, although their
strength varies as they propagate. This (modulated) travelling wave behaviour
for the field and flows can be clearly seen in Figure~\ref{fig:f05}{\sl b}.

\section{Discussion}
In this Letter we have identified a new, inherently nonlinear, mechanism for
the formation of coherent magnetic structures from a layer of weaker magnetic
field. The initial development of the magnetic buoyancy instability drives
flows that act so as to form isolated concentrations of magnetic flux. This
nonlinear state persists because the instability is continually driven from
the boundaries and saturation of the instability occurs via the net
redistribution of flux so as to remove average magnetic field gradients from
the interior of the computational domain. An intriguing feature of the flux
concentrations is that, once established, they travel --- taking the form of a
modulated wave.

The processes in the deep interior of the Sun that lead to the formation of
active regions are still poorly understood. Even the most optimistic estimates
from dynamo theory suggest that an upper bound for the large-scale
dynamo-generated field is of the order of $10^4$ G \citep[see,
e.g.][]{2003A&ARv..11..287O}. The energy of such a field is small, however, in
comparison with that of the strong downflows in the convection zone and, such
a field might be expected to be pinned down by the convection \citep[see,
e.g.][]{1998ApJ...502L.177T}. Clearly, therefore, some process must be acting
so as to create strong localised field structures from the dynamo field that
are capable of traversing the convection zone in order to form active regions.
Our calculations, although idealized, suggest a new mechanism that may play a
key role in this process.

We conclude by noting that the discovery of the mechanism arose from
considering a system where the instability is continually driven, in contrast
to nearly all previous studies in which the initial magnetic buoyancy
instability was investigated by means of run-down calculations. Both types of
experiment may be of importance for understanding the dynamics of the solar
interior. The magnetic buoyancy instability is a fast process that might be
captured by run-down calculations (which usually lead to the formation of
arched structures). On the other hand, the replenishment of toroidal field by
the dynamo should lead to a continual source of instability of field at the
base of the solar convection zone.

\acknowledgments The authors acknowledge financial support from PPARC, under
grant PPA/G/O/2002/00014, and the Leverhulme Trust. We are grateful to
N.H.~Brummell for helpful discussions and numerical advice.

\bibliography{apj-jour,ms}

\end{document}